\documentclass[a4paper,12pt]{article}
\usepackage{graphicx}

\usepackage{braket}
\usepackage{listings}
\usepackage{amsmath}
\usepackage{amssymb}
\usepackage{bussproofs}
\usepackage[toc,page]{appendix}
\usepackage{float}
\usepackage{mdframed}
\usepackage{mathtools}
\usepackage{url}
\usepackage[table,xcdraw]{xcolor}
\usepackage{amssymb}

\date{}

\def\limp {\mathbin{{-}\mkern-3.5mu{\circ}}}

\begin{document}

\title{Qumin, a minimalist quantum programming language}

\author{Alexander Singh$^{*}$        \and
        Konstantinos Giannakis \and Theodore Andronikos\\ \small{Ionian University, Department of Informatics}\\ \small{7 Tsirigoti Square, Corfu, Greece} \\ \small{\{p13sing, kgiann, andronikos\}@ionio.gr} \\ \small{* Corresponding author}} 

\maketitle

\begin{abstract}
In this work we introduce Qumin, a novel quantum programming language with a focus on providing an easy to use, minimalist, high-level, and easily extensible platform for quantum programming.
Qumin's design concentrates on encompassing the various interactions between classical and quantum computation via the use of two sublanguages: an untyped one that handles classical preparation and control, and one linearly typed that explicitly handles quantum routines. This allows the classical part of the language to be freely used for general programming while placing restrictions on the quantum part that enforce rules of quantum computing like the no-cloning of qubits.

We describe both the language's theoretical foundations in terms of lambda calculi and linear type systems, and more practical matters such as implementations of algorithms and useful programming tools like matrix and oracle generators that streamline the interaction of the
classical and quantum fragments of a program. 
Finally, we provide an experimental open-source implementation of an interpreter, typechecker and related tools for the language (which can be found in \url{https://github.com/wintershammer/QImp}).

\end{abstract}

\section{Introduction}

Since the conception of quantum computation in the later part of the last century, much work has been done in the development, design and understanding of algorithms that harness the potential of a quantum computer to outperform its classical analog. The existence of ``killer apps'' like Shor's and Grover's algorithms have spurred researchers to build models that simplify and streamline the process of creating algorithms that harness the unique capabilities of a quantum computer. Models like that of quantum circuits provide a low-level, hardware-like description of computation that although universal in its capabilities, might not be all that useful in high-level description methods that are appropriate for the analysis and implementation of algorithms.

Just as programmers of early computers turned to programming languages to bridge the gap between the low-level interface of the machine and the high-level conception of its operator, so too has research in quantum computing embarked in a search for languages that would fill the place of their classical counterparts, more so motivated by the need to come in terms with, and understand, the often times unintuitive, to our classical understanding, workings of quantum computers. Models like symmetric monoidal categories and their internal linear logic systems illuminate aspects that make quantum computers unique and give us a formalised, uniform and abstract/high-level way to understand them.

Qumin is a quantum programming language developed with the purpose of providing an accessible, minimalist, and easily modifiable programming environment, suitable for teaching, showcasing, and experimenting with algorithms in the simplest and most bare-bones way possible, while still maintaining the full power of abstraction that a language provides over models like quantum circuits. The driving force behind its design was creating a simple, yet theoretically sound, language that would
enable both newcomers and veterans of the field to experiment with, modify, and extend it to suit their specific needs. 

The design of Qumin is purposefully close to that of a classical functional language, with the common features serving as a familiar context for the classical programmer to approach quantum computing. It is the nature of Qumin's design to encompass the interaction between the classical and the quantum, as exemplified in the QRAM \cite{knill1996conventions} model and the slogan ``classical control, quantum data''. 

Qumin is composed of two sublanguages:
\begin{itemize}
\item An untyped one, that is suited for the classical portion of a program, in that it, for example, allows for unbound recursion. The fragments of a program described in it are used for preperatory/initialisation purposes (such as defining the initial states and operators to be used, and the number of iterations to run a program for), as well as for handling the execution of the program. Throughout the execution of a program, calls are made to quantum routines, defined in the sublanguage described below, that carry out the quantum parts of the program. 
\item A linearly typed one (based on multiplicative linear logic) that is restricted to functions that are formally correct to run on a quantum computer. Fragments of a program written in this sublanguage define the actual quantum computational routines.
\end{itemize}
Quantum computers are still experimental, noisy and expensive machines which only adds to the peculiarities of quantum information that make debugging as envisioned by classical programmers very difficult. Observing and tampering with quantum data modifies their state and so makes it difficult to inspect programs during their execution. Instead we would like to have a sort of certificate that our program behaves correctly and in
accordance with unique and often unintuitive rules like no-cloning and its dual no-deleting of qubits.

Such a certificate is provided by a rigid type system like that of linearly typed lambda calculi, on which the quantum fragment of our language is based. This however is not the only use of a type system. Type systems serve to illuminate and clarify what a program does, often serving as sort of a first-level documentation for a programmer. They also allow for complex data structures to be encoded in a uniform way and give powerful tools for programmers to define their own structures and data types that will suit their needs.

The main contributions of this work are:
\begin{itemize}
\item The description and implementation of a novel, easy to use, and easily extensible, programming language with two sublanguages: a classical sublanguage that is based on an extension of the untyped lambda calculus and a quantum one that is based on a fragment of linear type theory with multiplicative conjunction and exponentials, as described above.
\item Tutorial implementations of various quantum algorithms in Qumin; with a focus on the description of some useful patterns for quantum programming, like the matrix and oracle generators: functions that allow us to describe quantum programs in a clean, functional style
by writing linear operators as functions in a general setting, while still making use of matrices for efficient computations. These patterns also serve to streamline the interaction of classical and quantum computation as embodied by the two fragments of Qumin.
\end{itemize}
We believe that by designing the language to contain a familiar classical part and an easy-to-understand part dedicated solely to quantum routines, combined with the useful patterns and helper functions we discuss, allow Qumin to provide a user-friendly environment in which beginners can easily implement, and experiment with, quantum algorithms.

Qumin's classical fragment was recently \cite{singh2016genedis} used as a tool to showcase a preliminary version of some of the programming methods such oracles and matrix generators described here.

The following chapters describe the theoretical and practical foundations of Qumin. We start with a discussion of a ``naive" extension of the untyped lambda calculus with quantum primitives and its shortcomings in describing quantum computation.

We then describe a type system based on a fragment of linear logic which better suits the nature of quantum computing, in that it does not allow forbidden operations like cloning or deleting to take place and ensures proper usage of resources like qubits.

Afterwards, we present a showcase of Qumin's features by way of demonstration, providing implementations of QFT, Deutsch's and Grover's algorithms and variations thereof, as a way to explore the various design patterns and helpful routines that go into programming in a quantum context.

\section{Related Work}

Works dealing with the design of quantum programming languages like \cite{knill1996conventions,bettelli2003toward,selinger2004towards} have discussed some basic requirements one would
expect a quantum programming language to fulfill. These vary accordingly to the underlying paradigm,
with frequent requirements amongst other being: completeness, extensibility,
abstracting away and being independent from the underlying machinery and being expressive enough to allow one to
defne quantum data structures, oracles, handling of measurement, and handling
of quantum memory/registers.

Much work has been done in reformulating (finite) quantum mechanics and computation in the language of category theory \cite{abramsky2004categorical,selinger2007dagger,abramsky2009categorical,baez2010physics,heunen2012lectures}. Central to this area is the idea that variants of monoidal categories, which encapsulate ideas like the basic structure of composite systems in a resource-sensitive fashion, are a natural abstract model for quantum computation. The structure of such categories serve to illuminate aspects that are unique to quantum computers and help us understand the ``internal logic'' of quantum computation. Such categories allow us to reason about quantum operations in a high level of abstraction, making use of internal logics/type systems \cite{selinger2006lambda,selinger2009quantum,duncan2006types} and graphical methods such as string diagrams \cite{coecke2005kindergarten,selinger2010survey}. The aforementioned internal logics/type systems include the linear type system we describe in the following sections, which serves as the type system for Qumin's quantum fragment.

There already exists a number of quantum programming languages, both of imperative and functional styles, such as QML, Quipper, and QPL.

QPL, a functional programming language with well-defined operational semantics in terms of complete partial orders and superoperators first appeared in \cite{selinger2004towards}, and was a big influence in the development of other functional quantum programming languages.

QML \cite{altenkirch2005functional} is a functional quantum programming language, whose semantics are defined in terms of a category of Finite Quantum Computations, and
allow QML to be described in terms of superoperators and be translated to quantum circuits. QML's central features are built around the idea of controlling decoherence, a strict linear type system and quantum control in addition to quantum data. Interestingly QML's type system allows for contraction by way of sharing - not cloning qubits. 

Quipper \cite{green2013quipper} is focused on scalable, expressive and correct circuit creation and manipulation, providing many helpful features such as extensible data types, facilities for programming oracles, operating on and transforming circuits, describing subroutines and blocks. Quipper lacks a static linear type system (although it is a planned feature - as of \cite{green2013quipper}), choosing instead to check properties of the circuits at run-time.

There also exist a number of imperative languages. To start with, we have the aforementioned works by Knill which includes an imperative-style pseudocode and Bettelli et al. with a full scheme and implementation in C++. 
One of the first attempts at writing a full imperative language, called QCL, is described by \"{O}mer in \cite{omer1998procedural}, with a C-like syntax and high-level quantum programming features such as automatic memory management, user-defined operators, and automatic derivations of them. It also includes a classical sublanguage.
Sanders and Zuliani \cite{sanders2000quantum} have also proposed an imperative language, qGCL, which is based on a guarded command language and enjoys rigorous semantics and an associated refinement calculus.
Another imperative language, LanQ, was introduced by Mlna\v{r}\'{i}k\cite{mlnarik2007operational}, who also provided operational semantics for it, as well as a type soundness proof for the non-communicating part of the language.

Imperative languages, as noted in \cite{selinger2004brief}, frequently lack type systems comparable to those of their functional counterparts, which are sophisticated enough to allow for complete checking to happen statically.
The approach we have taken in designing Qumin is more closely related to that of other quantum functional languages, rather than the aforementioned imperative languages. Instead of viewing programs as steps to be executed, we describe them as a composition of well-defined easily modifiable functions/routines of both classical and quantum nature. 
This approach allows use to use high-order functions such as matrix and oracle generators that operate on other functions to produce useful components such as matrix representations and unitary oracles from classical functions. We also rely on a static type system to ensure proper handling of quantum resources such as qubits.

A more recent example of a domain-specific language can be found in \cite{wecker2014liqui}. LIQUi$\ket{}$'s language design focuses on a statically type-checked functional language with an isolated physical model, eschewing the use of linear types to allow instead for the manipulation of qubits in mutable ways not allowed by linear types.

For the sake of convenience, we also present some of the above discussed languages in tabular form in Table~\ref{table-languages}.

A more comprehensive survey of programming language research can be found in the aforementioned work by Selinger \cite{selinger2004brief}, as well as the works of Gay\cite{gay2006quantum} and Sofge\cite{sofge2008survey}.

\begin{table}[]

\caption{A selection of some quantum programming languages.}
\label{table-languages}
\begin{tabular}{ll>{\footnotesize}l}
\hline
\multicolumn{1}{|l|}{Name} & \multicolumn{1}{l|}{Style} & \multicolumn{1}{l|}{Notes}                                                                                      \\ \hline
QCL                        & Imperative                 & \multicolumn{1}{m{0.8\textwidth}}{Has classical sublanguage, multiple high-level programming features.}\\
qGCL                       & Imperative                 & \multicolumn{1}{m{0.8\textwidth}}{Emphasis on algorithm derivation and verification.}\\
LanQ                       & Imperative                 & \multicolumn{1}{m{0.8\textwidth}}{Full operational semantics, proven type soundness.}\\
Quipper                    & Functional                 & \multicolumn{1}{m{0.8\textwidth}}{Focus on scalability, plans to include linear types for static checks (currently done at run-time).}\\
QPL                        & Functional                 & \multicolumn{1}{m{0.8\textwidth}}{Statically typed, denotational semantics in terms of CPOs of superoperators.}\\
QML                        & Functional                 & \multicolumn{1}{m{0.8\textwidth}}{Linearly typed, focused on weakening - not contraction. Quantum control and quantum data.}\\
\rowcolor[HTML]{EFEFEF} 
Qumin                      & Functional                 & \multicolumn{1}{m{0.8\textwidth}}{Two sublanguages (untyped and linearly typed). Focus on ease of use and clean, functional style of programming.}
\end{tabular}
\end{table}

\section{The foundations of Qumin}
\subsection{Extending the untyped lambda calculus with quantum primitives - The $\lambda_{H}$ system}

Before delving into the type system that the quantum fragment of Qumin is based on, let us first study a variation of the untyped lambda calculus in preparation for the following chapters. The basic ideas from this extension are used by the classical part of the language to handle the definitions, preparation and initialisation of quantum data that will be fed to the quantum fragment for execution. We extended the untyped lambda calculus with primitive operations and constants, in order to facilitate operations in  Hilbert spaces $H$, in direct accordance with postulates of quantum mechanics, as described in \cite{nielsen2010quantum}.

\begin{figure}[H]
\begin{center}
	t := \hfill (term) \\
	\qquad $x$ \hfill (variable) \\
	\qquad $v$ \hfill (vector) \\
	\qquad $U$ \hfill (operator) \\
	\qquad $(U \cdot v)$ \hfill (operator application) \\
	\qquad $(v \otimes v)$ \hfill (tensor product) \\
	\qquad $measure(v)$ \hfill (measurement) \\
	\qquad $\lambda x.t$ \hfill (abstraction) \\
	\qquad $t \ t$ \hfill (application) \\
\end{center}

\end{figure}
\noindent Where, for a given Hilbert space $H$,
\begin{itemize}
\item $[\![v]\!]$ belongs to the set of normalized vectors of $H$.
\item $[\![U]\!]$ belongs to the set of matrix representations of unitary operators of $H$. 
\item $[\![U \cdot v]\!]$ is operator application, by way of matrix multiplication.
\item $[\![v \otimes v]\!]$ is the tensor/Kronecker product of two vectors.
\item $[\![measure(v)]\!]$ is measurement of state $v$ in the computational basis, returning the state after collapse.
\end{itemize}

We can see then, that the addition of normalised vectors corresponds to the first postulate, the addition of operators and operator application correspond to the second postulate, the addition of the measurement operation corresponds to the third postulate
and finally the addition of the tensor product operation corresponds to the forth one.
In practice, the parenthesis and the multiplication dot can be omitted when the meaning is clear.
\\For example, Deutsch's algorithm, assuming $H$ corresponds to the Hadamard transformation matrix and $I$ to the identity matrix, is expressed in $\lambda_H$ as such:

\centerline{$\lambda U_f.measure((H \otimes I)U_f(H \otimes H)(\ket{0} \otimes \ket{1}))$}

\noindent where the input $U_f$ is the matrix that corresponds to the oracle of a binary function $f:\{0,1\}\to\{0,1\}$:\\
\centerline{ $U_f(\ket{x,y})= \ket{x,y \oplus f(x)}$ }\\
where $\oplus$ is the exclusive or/modulo-2 addition operation.

\subsection{The $\lambda_{H\limp}$ system}

While $\lambda_H$ is capable of describing quantum computation, it unfortunately also allows for statements that are nonsensical or even plain wrong in the quantum-computing context.
For example $\lambda qubit.qubit \otimes qubit$ is perfectly valid as a $\lambda_H$ expression, but its clearly wrong in a very concrete sense: it allows us, for a given qubit, to construct a bipartite system of it with itself.
This of course is not possible for arbitrary states, as exemplified by the no-cloning theorem. 

It would be helpful then to somehow restrict the set of expressions to those that we consider to be a faithful description of quantum computing.
To this end we introduce a type system for $\lambda_H$, based on a fragment of linear logic, a good exposition of which one can find in \cite{girard1995linear}. Assuming linear contexts $\Delta$ and $\Delta'$, as well as a regular unrestricted context $\Gamma$, the type system's rules can be presented as in Figure~\ref{fig:rules}, while the types and terms of the calculus are presented in Figure~\ref{fig:tts}, where ``$prim$'' includes primitives such as integers, qubits and a family of functions as seen in Figure~\ref{fig:prims}.

Semantically, it is convenient to think of the the quantum, multiplicative part of the type system in terms of an appropriate (closed symmetric monoidal) category of finite Hilbert spaces:

\begin{itemize}
\item $qubit$ corresponds to the two-dimensional Hilbert space which describes qubits,
\item $X \otimes Y$ corresponds to the monoidal product of two objects $X$ and $Y$,
\item $X \limp Y$ corresponds to the internal hom object $X \limp Y$.
\end{itemize}

The semantics for the exponential $!$-modality are quite more complex, but we can informally think of values of type $!A$ as ``classical'' or ``re-usable'' values of type $A$.

As mentioned before, further and rigorous discussions of semantics for (multiplicative) linear logic/linear type systems, as well as their connection to monoidal categories, can be found in the works such as \cite{selinger2006lambda,selinger2009quantum,duncan2006types}.

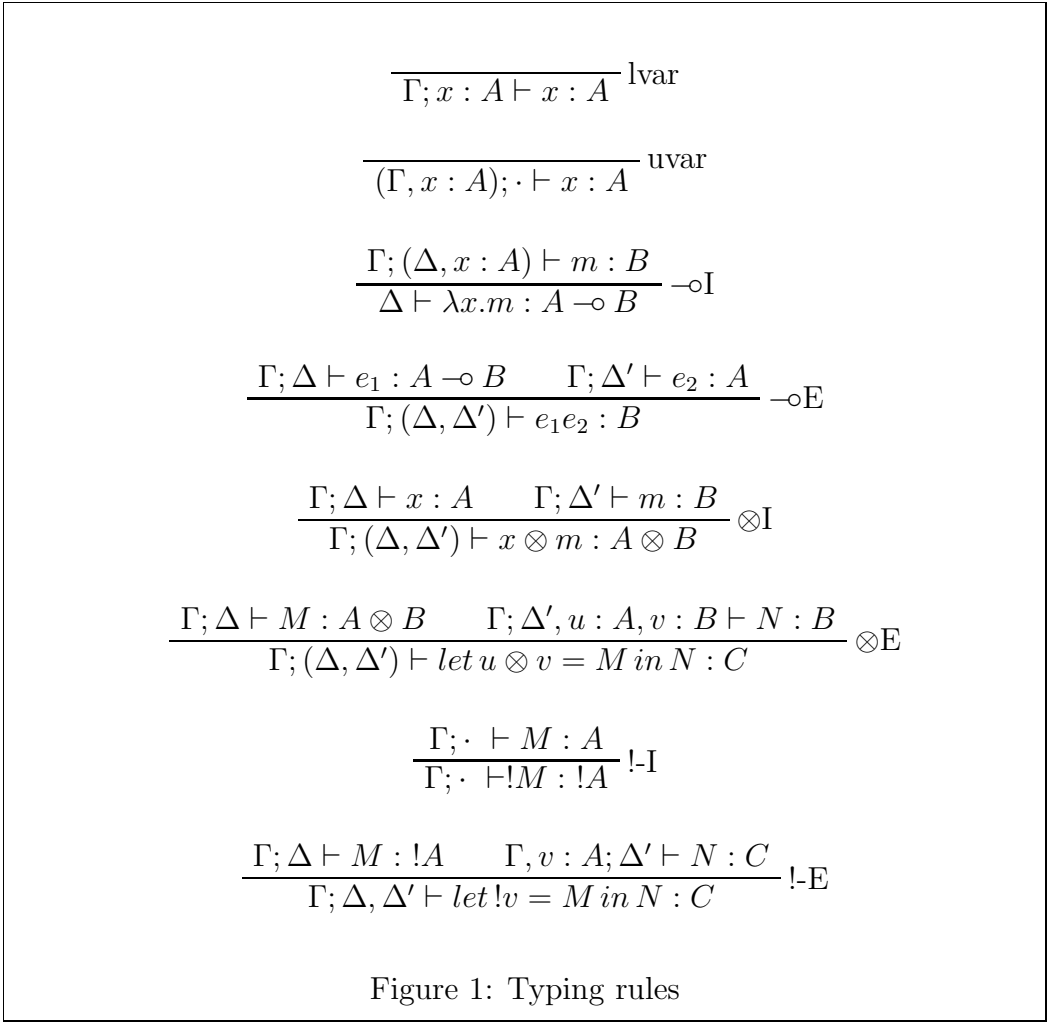
\begin{mdframed}
\begin{figure}[H]

\begin{prooftree}
\AxiomC{}
\RightLabel{lvar}
\UnaryInfC{$\Gamma; x : A \vdash x : A$}
\end{prooftree}

\begin{prooftree}
\AxiomC{}
\RightLabel{uvar}
\UnaryInfC{$(\Gamma, x : A); \cdot \vdash x : A$}
\end{prooftree}

\begin{prooftree}
\AxiomC{$\Gamma; (\Delta, x : A) \vdash m : B$}
\RightLabel{$\limp$I}
\UnaryInfC{$\Delta \vdash \lambda x.m : A \limp B$}
\end{prooftree}

\begin{prooftree}
\AxiomC{$\Gamma; \Delta \vdash e_1 : A \limp B$}
\AxiomC{$\Gamma; \Delta' \vdash e_2 : A $}
\RightLabel{$\limp$E}
\BinaryInfC{$\Gamma; (\Delta,\Delta') \vdash e_1 e_2 : B$}
\end{prooftree}

\begin{prooftree}
\AxiomC{$\Gamma; \Delta \vdash x : A$}
\AxiomC{$\Gamma; \Delta' \vdash m : B $}
\RightLabel{$\otimes$I}
\BinaryInfC{$\Gamma; (\Delta,\Delta') \vdash x \otimes m : A \otimes B$}
\end{prooftree}

\begin{prooftree}
\AxiomC{$\Gamma; \Delta \vdash M : A \otimes B$}
\AxiomC{$\Gamma; \Delta', u : A,  v : B \vdash N : B $}
\RightLabel{$\otimes$E}
\BinaryInfC{$\Gamma; (\Delta,\Delta') \vdash let \, u \otimes v = M \, in \, N :  C$}
\end{prooftree}

\begin{prooftree}
\AxiomC{$\Gamma; \cdot \ \vdash M : A$}
\RightLabel{$!$-I}
\UnaryInfC{$\Gamma;  \cdot \ \vdash !M :\ !A$}
\end{prooftree}

\begin{prooftree}
\AxiomC{$\Gamma; \Delta \vdash M : \ !A$}
\AxiomC{$\Gamma, v : A; \Delta' \vdash N : C$}
\RightLabel{$!$-E}
\BinaryInfC{$\Gamma; \Delta,\Delta' \vdash let \, !v = M \, in \, N :  C$}
\end{prooftree}

\caption{Typing rules}
\label{fig:rules}
\end{figure}
\end{mdframed}

The rules presented in Figure~\ref{fig:rules} however, as currently formulated, pose a certain critical problem for practical typechecking, in regards to how the splitting of contexts ($\Delta,\Delta'$) is to take place.
To this end, the rules used in the implementation of our interpreter use a method of splitting the context where each premise ``consumes'' a part of the context and passes the rest on,  in the tradition of \cite{Pfenning}.

\begin{mdframed}
\begin{figure}[H]
$Type \; U,V =  qubit \ | \ int \ | \ list \ |\ U \limp V \ | \  U \otimes V \   | \ !U$ \\ 

$Term \; t,u =  \ prim \ | \ var \ | \ \lambda var . t  \ | \ t u  \ | \ let \; x \otimes y = t  \; in \; u$\\
\caption{Types and terms}
\label{fig:tts}
\end{figure} 
\end{mdframed}

\begin{mdframed}
\begin{figure}[H]
$\cdot : (qubit^{\otimes n} \limp qubit^{\otimes n}) \limp qubit^{\otimes n} \limp qubit^{\otimes n}$, \\
$\otimes : qubit \limp qubit \limp qubit \otimes qubit$, \\
$applyN : (qubit^{\otimes n} \limp qubit^{\otimes n}) \limp qubit^{\otimes n} \limp int \limp qubit^{\otimes n}$\\
$measure : qubit^{\otimes n} \limp \; !qubit^{\otimes n}$, \\
$tensorOp : (qubit^{\otimes n} \limp qubit^{\otimes n}) \limp (qubit^{\otimes m} \limp qubit^{\otimes m}) \limp (qubit^{\otimes n} \otimes qubit^{\otimes m} \limp qubit^{\otimes n} \otimes qubit^{\otimes m})$\\
$subsystems : qubit^{\otimes n} \limp \; list \limp \; !qubit^{\otimes n}$\\
\caption{Primitive function signature schemes, for $n,m \in \mathbb{N}$. }
\label{fig:prims}
\end{figure}
\end{mdframed}

Note that $apply/\cdot$ is essentially just explicit function application. While Qumin's typechecker has no problem checking either \texttt{apply(U,v)} or \texttt{U(v)}, we present it here to streamline the transition from the untyped lambda calculus presented in the last chapter, where $apply$ had the actual function of specifying how to reduce expressions of the form $U \cdot v$, via matrix/vector multiplication. This syntax is also useful in compositions such as \texttt{apply(tensorOp(H,H),[1 0])}, as we'll see later on.

Throught this work we, for the sake of clarity and readability, will make use of standard shorthand notations such as:
\begin{itemize}
\item $\lambda x_1...x_n.E = \lambda x_1.\lambda x_2. ...\lambda x_n.E$
\item $\lambda x_1 \otimes x_2.E = \lambda y. ($let $ x_1 \otimes x_2  = y$ in $E)$
\end{itemize}
\subsection{Connections with other models of computation}

As mentioned before, a popular way to model quantum computation is that of the quantum circuits. Quantum
circuits model computation as a sequence of quantum gates and measurements applied to a
quantum register. Again a clear connection can be made with the aforementioned four postulates.
 The input quantum registers are composite systems that correspond to the tensor product of multiple qubits, quantum gates correspond to unitary operators
and measurements of course correspond to measuring the registers in the computational basis.

\textit{Definition.} A quantum circuit describes the evolution of an initial quantum register by the sequential application of quantum gates and finally the measurement of such a register in the computational basis.
More concretely, a quantum circuit starts with a prepared initial state $\ket{i} = \ket{i_1} \otimes \ket{i_2} \otimes ... \otimes \ket{i_n}$, to which at every instant $t$
we apply a list of quantum gates (including the identity gates which are usually not explicitly shown). Such a collection $U_k = (U_1...U_k)$ can be written as a single unitary operator 
$U_t = U_1 \otimes ... \otimes U_k$. Then the full unitary evolution of such a system is written as such: $U_c = U_t * U_{t-1} * ... U_2 * U_1$.

\textit{Definition.} A ``routine'' is a $\lambda_{H\limp}$ expression of the form:\\

$\lambda \ket{initial} : qubit^{\otimes n} \; U1 : qubit^{\otimes n}\ \limp qubit^{\otimes n} \; ... \; U_n : ~\\
qubit^{\otimes n} \limp qubit^{\otimes n}.measure(U_1, U_2 ... U_n \ket{initial})$~\\ \\
for some natural n.\\

\noindent\textbf{Proposition}: Quantum circuits and $\lambda_H\limp$ are equivalent descriptions of finite quantum computation.\\

It is clear that the two models encapsulate the same three notions of an initial state, a sequence of unitary operations and a final measurement.
Explicitly, let us a finite quantum computation to be a procedure where, starting with an initial state $\ket{\psi} \in H$, we apply a sequence of unitary operators $U = U_1, U_2, U_3 , ... , U_n$
before finally doing a measurement in the computational basis. Then the circuit that describes such a computation is one with $\ket{\psi}$ as an initial configuration and for each instant
$t$ the operator that is applied to the system is the $i-th$ operator of the sequence U.
The equivalent  $\lambda_{ H \limp}$ expression is the above defined routine, applied to $\ket{\psi}, U_1, U_2, ... , U_n$ where $U_i \in prim$.

This means that we are able to trivially translate standard quantum circuits to programs in our language, allowing easy implementation of, and experimentation with, well-known quantum algorithms.

\section{Introducing Qumin}

We have already seen that $\lambda_{H \limp}$ is incapable of recursion, which unfortunately severely restricts its use in general purpose programming. It is however extremely useful for checking the correctness of, and mathematically reasoning about, our actual quantum operations.
With this in mind, Qumin makes use of both $\lambda H$ and $\lambda H \limp$. This means it has both an untyped component and a strictly typed component. The untyped component handles all general definitions, preparations, initialisation and classical control, while the linearly typed one handles explicit quantum routines, that would be executed on the actual ``quantum component'' of a quantum computer.
We provide an experimental implementation of an interpreter for Qumin, which can be found in \cite{qimp} (which also contains links to a more beginner-friendly tutorial of Qumin).

\subsection{The design of Qumin's interpreter}

As discussed previously, Qumin's design is based on the use of two sublanguages, each describing a corresponding fragment of a program:

\begin{itemize}
\item A classical one to handle preparation and definitions, as well as execution of programs
\item A quantum one to handle the definition, and verify correctness of, quantum functions/routines.
\end{itemize}

The typical interpretation process of a Qumin program is as follows: 
\begin{itemize}
\item Typecheck and load libraries defined in the quantum fragment. This marks the ``static'' part of the interpretation process - any errors in quantum routines immediately end the interpretation with a relevant error message.
\item Load libraries defined in the classical fragment.
\item Execute the program, as described by the classical fragment. Any time a function defined in the quantum fragment is called, generate argument constraints to check against the type signature. Doing this, we can make sure that the arguments supplied by the classical fragment match the specifications defined by the quantum one. 
\end{itemize}

Currently, the interpreter carries out all computations internally, using Python primitives and the \texttt{numpy} library, but it would be interesting to consider cross-compilation and interfacing with current frameworks for quantum computation such as \cite{steiger2016projectq} and \cite{smith2016practical}.

An experimental implementation of said interpreter can be found in \cite{qimp}.

\subsection{Classical programming in Qumin}

We start our showcase of Qumin with a description of its classical aspects.

\subsubsection{Primitive datatypes and variables.}
The primitive datatypes of the classical part of Qumin are:

\begin{itemize}
\item{Integers: \texttt{5, -32}}
\item{Floats:  \texttt{0.2, -99.212}    }
\item{Complex numbers: \texttt{1+5i, -3.2-5i}}
\item{Booleans: \texttt{\#t,\#f}}
\item{Strings: \texttt{``Hello world!"}}
\item{Lists:  \texttt{[1 2 3],[[1 2] [3 4]]}}
\end{itemize}

Variables, used in the mathematical sense of the world (identifiers referring  to immutable values), are defined as such:
\begin{lstlisting}
let identifier = expression
\end{lstlisting}

For example:

\begin{lstlisting}
let myString = "Hello world!"
let six = (3 + 3)
\end{lstlisting}

One can refer to the grammar of Qumin, presented in the appendix, for specifics about what constitutes a valid identifier or expression.

\subsubsection{Functions}

As mentioned in the chapters before, Qumin is based on lambda calculus/linear type theory, where the notion of a function is captured by lambda abstractions.
For example $\lambda x.x+5$ is written in Qumin as such:

\begin{lstlisting}[]
lambda(x){
  (x + 5)
}
\end{lstlisting}

Lambda abstractions can be invoked in-line by including arguments in a parenthesis as such:

\begin{lstlisting}[]
lambda(x,y){
  (x + y)
}(3,5)
\end{lstlisting}
which would evaluate to \texttt{8}.

Qumin, being a functional programming language, places great significance in the notion of functions as the building blocks of a program.
Functions are first-class citizens, in that they can be passed around and returned as any other primitive, like lists or numbers, and can be bound to identifiers. 
The returned value of a function is the last evaluated expression in its body.
For example, a function that takes another function and applies it to an argument:

\begin{lstlisting}
lambda(f,x){ 
  f(x)
}(lambda(x){(x + x)},5)
\end{lstlisting}
which evaluates to \texttt{10}.

While anonymous functions are theoretically sufficient for computation, we prefer their named counterparts for the sake of convenience.
To define a named function, we attach a lambda abstraction to an identifier.
For example $f(x) = x + 5$ is written in Qumin as such:

\begin{lstlisting}[]
let f = lambda(x){
  (x + 5)
}
\end{lstlisting}

\noindent And can be invoked as such:

\begin{center}
\texttt{f(5)\\}
\end{center}
which of course evaluates to 10.\\

Qumin supports implicit partial application:
\begin{lstlisting}[]
let f(x,y){
  (x + y)
}
let partiallyApplied = f(10)
partiallyApplied(30)
\end{lstlisting}
\noindent What happened here is that, by only supplying one of the two expected arguments, we were given back a new function that looks like this:\\

\begin{lstlisting}
lambda(y){
  (10 + y)
}
\end{lstlisting}
which we then proceed to bind to the identifier ``\texttt{partiallyApplied}'' and then invoke with argument 30.
This last call would evaluate to 40.

Finally, specifically in the case of binary functions, we can also call them in infix notation: \texttt{(argument1 function argument2)}.
For example:
\begin{lstlisting}[]
let myOp = lambda(x,y){
  (x + (3 * y))
}
(5 myOp 10) => 35
\end{lstlisting}
This is the actual reason we have to surround every arithmetic expression in parenthesis.
Arithmetic operators (\texttt{+,-,*/}) in Qumin are defined as any other function would be, we just call them infix for clarity.
We could just as well invoke them as such:\\

\begin{lstlisting}[]
+(3,5) => 8
-(10,-3) => 13
\end{lstlisting}

\subsection{Quantum programming in Qumin}

\subsubsection{Vectors and Matrices}

Vectors and matrices are of central importance in quantum computing, where they represent the state/qubits of a system and unitary operators/gates respectively.
In Qumin, vectors and matrices are implemented using lists and lists of lists.

For example a state $\ket{\psi} = a\ket{0} + b\ket{1}$ in the two-dimensional space $H$, is written in Qumin as such:\\

\begin{center}
\texttt{let psi = [a b]}
\end{center}

While, for example, the identity matrix that corresponds to the identity operator in $H$ would be written as:

\begin{lstlisting}[]
let identity = [[1 0]
                [0 1]]
\end{lstlisting}

Of course, as the dimension of $H$ increases, the process of writing matrices by hand quickly gets unwieldy.
For example, for a space of 4 qubits one would be expected to write a 16x16 (256 values) matrix by hand.
To combat this problem, we could eschew the use of matrix representations and work with linear operators as functions.
This alleviates the aforementioned problem of having to manually define multi-dimensional matrices by hand. 

For example, the identity operator is always $f(x) = x$, regardless of the dimension of the space.
Unfortunately this also has the side-effect of making things like checking if an operator is unitary or finding eigenvalues/eigenvectors and hermitian adjoints of operators difficult, 
while also introducing slowdowns in computations. 

\subsubsection{Matrix Generators}

The solution to the aforementioned dilemma is given by functions called matrix generators.
Matrix generators allow us to make use of an operator in its function form where convenient and in its matrix form otherwise.
A generator is a function that when given an linear operator $f: H \to H$ and a basis $\{v_i\}$ of H, it generates $f$'s matrix representation on $H$ with respect to the basis.
This allows us to write linear operators as functions, composing them and manipulating them as one would expect to manipulate a mathematical operator,
and when we want to make use of its matrix representation, all we have to do is invoke the generator on it.\\

\begin{mdframed}
\noindent\textbf{Matrix Generator Algorithm.} \\
\textbf{Inputs:} $f: H \to H, \{v_i\}$\\
\textbf{Outputs:} $M_{dim(H) \times dim(H)}$ \\
\textbf{0:} M $\leftarrow$ [ ] \\
\textbf{1:} For \textit{v} in  $\{v_i\}$: \\
\textbf{2:} \qquad append $f(v)$ to M \\
\textbf{2:} transpose M, making $f(v)$s its collumns
\end{mdframed}

For example, the identity operator is defined as such:

\begin{lstlisting}[]
let identity = lambda(vec){
  vec
}
\end{lstlisting}

Then generating, for example, the identity matrix on a 16-dimensional (4-qubit) Hilbert space, amounts to running:

\begin{lstlisting}[]
generateMatrix(identity,16)
\end{lstlisting}

Apart from allowing us to avoid writing big matrices by hand, generators allow us to define operators in a mathematical, easily-understood, and general with respect to dimension, way.
For example the Quantum Fourier Transform is written in Qumin as such:\\

\begin{lstlisting}[]

--load generator

let omega = lambda(jj,k,N){
  exp((fold(*, [2 pi 0+1i jj k]) / N))
}

let qfSum = lambda(limit,vec,index,N){
  if ((limit = 0)){
      0
  }
  else {
      ((omega(index,limit,N) * car(vec)) +
      qfSum((limit - 1), cdr(vec), index, N))
  }
} 



let outerMult = lambda(vec,index,N){
  if((N = index)){
    []
  }
  else{
    append(((1 / sqrt(N)) * qfSum(N,vec,index,N)),
           outerMult(vec,(index + 1),N))
  }

}

let qft = lambda(vec){
  let N = len(vec)
  outerMult(vec,0,N)
}

\end{lstlisting}

As we can see, the Qumin implementation closely follows the mathematical expression of QFT:
\[ y_k = \frac{1}{\sqrt{N}}  \sum\limits_{j=0}^{N - 1} x_j \omega^{jk} \]
Where:
\[ \omega^{jk} = e^{2 \pi i \frac{jk}{N}} \]

The function \texttt{omega} implements $\omega^{jk}$ (i.e the $N^{th}$ root of unity), \texttt{qfSum} implements the sum $ \sum\limits_{j=0}^{N - 1} x_j \omega^{jk}$,
and \texttt{outerMult} builds the transformed vector ($y_k$) by multiplying each result of \texttt{qfSum} by $\frac{1}{\sqrt{2}}$.

\subsubsection{The quantum fragment of Qumin}

We now introduce the quantum fragment of Qumin. As we'll see in the examples that follow, implementing a quantum algorithm in Qumin follows a specific pattern: we have a classical untyped part of the program that handles definitions and initialisation, that then calls a linearly typed function that implements the algorithm itself. These linearly typed functions are much more restricted than their classical counterparts - they are based on the use of the primitive functions discussed in section 5.1 and their structure is reminiscent of the $\lambda_{H\limp}$ routine defined in 5.2. 
These routines are defined in much the same way as other Qumin functions, with the difference that these include type signatures.

For example a simple routine that applies an operator $U$ to a qubit $q$ and then measures it:

\begin{lstlisting}[,mathescape=true]
let simpleRoutine = lambda(q : qubit, U : operator[1]){
  measure(
    apply(U,q))
}
\end{lstlisting}

This routine would then be loaded by a classical program, which contains the definitions/initialisation of the qubit and the operator. The classical program would then call the routine, passing the qubit and the operator as arguments.
During the static part of the program's execution, the typechecker would check the correctness of \texttt{simpleRoutine}, based on the rules described in section 5.

Once the routine itself has been checked for correctness, the Qumin would handle the interaction between the classical and quantum fragments by generating a list of constraints and specifications that the inputs, provided by the classical part, must follow. These constraints are based on the signatures described above. In this example, the first input passed should be a list that represents a single qubit, while the second input should be a 2x2 matrix (a list of lists) representing $U$. During the execution - that is, the dynamic part - any argument passed from the classical side to a routine is checked against these constraints.

Some of the types that can be used in these type signatures are:

\begin{itemize}
\item \texttt{qubit} : the type of qubits
\item \texttt{int} : the type of integers
\item \texttt{list} : the type of lists
\item \texttt{type1 * type2} : the multiplicative conjunction/tensor product of the types \texttt{type1} and \texttt{type2}
\item \texttt{type1 > type2} : the type of functions from \texttt{type1} to \texttt{type2}
\item \texttt{operator[n]} : shorthand for the type of operators acting on $n$ qubits
\item \texttt{!\{type\}} : denotes the !/exponential modality applied to \texttt{type}
\end{itemize}

\section{Examples of quantum algorithm implementations}
\subsection{Deutsch's Algorithm}

We will now proceed with an implementation of Deutsch's algorithm.
Once again, we look back to our theoretical foundations where computation is based on primitive operations like: $\cdot ,\otimes$, $measure$, $tensorOp$, which in Qumin are defined as functions named
$\cdot$, $\otimes$, \texttt{measure} and  \texttt{tensorOp} respectively. If one wishes to avoid using unicode, he can use the aliases \texttt{apply} for $\cdot$ and \texttt{tensor} for $\otimes$ instead.

For example, we already have presented Deutsch's algorithm in $\lambda_H$ and $\lambda_{H\limp}$ so let us present the Qumin version:\\

\begin{lstlisting}[,mathescape=true]
--qload deutschTypes
--load operators
--load generator




let H = generateMatrix(hadamard,2)
let I = generateMatrix(identity,2)

let fConstant = lambda(x){
  [1 0]
}

let fBalanced = lambda(x){
  x
}

let state = tensor([1 0],[0 1])

let Uf = oracle(generateMatrix(fConstant,2))

}
\end{lstlisting}

\begin{lstlisting}[,mathescape=true]
let deutschRoutine = lambda(state : qubit * qubit, H : !{operator[1]}, I : !{operator[1]}, U : !{operator[2]}){

  measure(
      apply(tensorOp(H,I),
      apply(U,
      apply(tensorOp(H,H),
      state))))
      
}
\end{lstlisting}

As we can see, the body of \texttt{deutschRoutine} closely resembles the corresponding lambda version:\\
$\lambda U_f : qubit \otimes qubit \limp qubit \otimes qubit.measure((H \otimes I)U_f(H \otimes H)(\ket{0} \otimes \ket{1}))$\\
(assuming H and I are primitives).

Running Deutsch's algorithm on the first example function, $f(x) = \ket{0}$ , gives us an output of:
\begin{lstlisting}[,mathescape=true]
=> deutsch(fConstant)

Probability of state 0 is 0.5
Probability of state 1 is 0.5
Probability of state 2 is 0.0
Probability of state 3 is 0.0
System collapsed to state: 0
\end{lstlisting}

This is the output message provided by the \texttt{measure} function. Since we will examine such outputs in the following examples too, it would be beneficial to take a minute and explain its structure.
The \texttt{measure} function performs a measurement on the computational basis, collapsing its input and returning a value of type $!A$ for an input of type $A$. Apart from return the aforementioned value, \texttt{measure} also prints a message, as seen above, which lists the probabilities we had of measuring each of the basis states post-measurement, as well as finally listing the state that our system collapsed to. The naming of the states follows the usual convention of labeling states using numbers from 0 to $n$ for an $n$-dimensional state.

\noindent While doing the same with the second example function, $f(x) = x$, gives us:
\begin{lstlisting}[,mathescape=true]
deutsch(fBalanced)

Probability of state 0 is 0.0
Probability of state 1 is 0.0
Probability of state 2 is 0.5
Probability of state 3 is 0.5
System collapsed to state: 3
\end{lstlisting}

\noindent As expected.\\

One may notice that in the implementation of Deutsch's algorithm we made use of a function called \texttt{oracle}.
The \texttt{oracle} function converts classical operators to unitary ones, allowing us to use them in our quantum computations. To do this, \texttt{oracle} expects as input
a function $f$ that operates on binary strings and creates a new operator $U$ that operates on a composite space, the tensor product of the domain of f seen as a qudit and an additional helper qudit. 
That is, for $f(x)$, \texttt{oracle} creates $U(x,y)$ defined as such: $U(x,y) = (x, y \oplus f(x))$, where $\oplus$ is the (bitwise) addition modulo 2/exclusive or operation.\\

\noindent\textbf{Proposition 2.} Matrices generated by \texttt{oracle} are unitary.\\
\textbf{Proof.} See Appendix A.

\subsection{Grover's algorithm}

Let us now discuss an implementation of Grover's algorithm, one of the more famous quantum algorithms. 
Grover's algorithm finds the specific input for which a binary function returns 1. For example 
let f be the function which maps database entries to 0 or 1, where $f(x)=1$ if and only if $x$ satisfies a search criterion.
Indeed this is the most discussed application of Grover's algorithm, namely database search.

The steps of Grover's algorithm are as follows:\\

\begin{enumerate}
\item Start with a uniform superposition, over all states
\item Repeat the Grover iteration O($\sqrt{N}$) times, where N is $dim(H)$.
\end{enumerate}

The Grover iteration is:

\begin{enumerate}
\item Apply the oracle operator: $U_f$
\item Apply the Grover diffusion operator: $H^{\otimes n} (2\ket{0}\bra{0} - I) H^{\otimes n}$
\end{enumerate}

Alternatively, the Grover diffusion operator can be written as: $ (2\ket{\psi}\bra{\psi} - I) $ for $\ket{\psi} = 1/\sqrt{N}  \sum\limits_{i=1}^{N-1} \ket{i}$

Let's start our implementation by focusing on a specific case: a search in a database of 4 elements. This is modeled using a 4-dimensional space H, each basis of which corresponds to an element of the database.

The first thing we must do is define the function $f$, that is the function that returns 1 for the element we are searching for and 0 for all else.
The straightforward way would be to define it as a function:
\begin{center}
\begin{lstlisting}[,mathescape=true]
let f = lambda(string){
  if((string = stringImSearchingFor)){
    [1 0]
   else{
    [0 1]
  }
}
\end{lstlisting}
\end{center}
But this would require us to manually write a new function for each search, which of course is not very practical.
Instead, to make our lives easier, let's define it as a function that given a string $s$, returns an appropriate (anonymous) function that works like the $f$ described above:

\begin{center}
\begin{lstlisting}[,mathescape=true]
let f = lambda(string){
  lambda(x){
    if((x = string)){
      [0 1]
    }
    else{
      [1 0]
    }
  }
}
\end{lstlisting}
\end{center}

Now we have to define the oracle operator for our function. But Qumin's \texttt{oracle} only accepts a matrix as input, so we firstly have to generate a matrix for our function.
Once again, we can make use of our generator function to do just that:

\begin{center}
\begin{lstlisting}[,mathescape=true]
Uf = oracle(generateMatrix(f(string),4))
\end{lstlisting}
\end{center}

We can now define the last component of our iteration, the Grover diffusion operator, by a straightforward translation of the mathematical formula:

\begin{center}
\begin{lstlisting}[,mathescape=true]
let groverOper =  ((2 * outer(average,average)) - I)
\end{lstlisting}
\end{center}

Let's bundle the last two steps to one function which we will call the \texttt{groverIter}, that takes as arguments: the state, the oracle, and a number of times which we want to repeat the iteration and calls the corresponding quantum routine:

\begin{center}
\begin{lstlisting}[,mathescape=true]
let groverIter = lambda(state,Uf,times){
  
  let average = apply(Htwo,[1 0 0 0])

  let groverOper =  ((2 * outer(average,average)) - I)

  let opb = tensor(groverOper,generateMatrix(identity,2))

  let iteration = apply(opb,Uf)
  
  groverRoutine(state,iteration,times)
}
\end{lstlisting}
\end{center}

\begin{center}
\begin{lstlisting}[,mathescape=true]
let groverRoutine = lambda(state : qubit * qubit * qubit, iteration : !{operator[3]}, times : int){
  measure(applyN(iteration,state,times))
}
\end{lstlisting}
\end{center}

Recall that the \texttt{oracle} function generates (the matrix of) an operator that takes $\ket{x, y}$ to $\ket{x, y \oplus f (x)}$, where y is an auxiliary qubit.
But we want our Grover diffusion operator G to only work on the database qubits and not the auxiliary qubit of the oracle, so we must define a new operator
$G\otimes I$ where $I$ is the 2-d identity operator, which applies the Grover diffusion to the ``top'' qubit register and leaves the ``bottom'' one unchanged.

We can now define the function that executes Grover's algorithm, searching for our database for a given string:

\begin{center}
\begin{lstlisting}[,mathescape=true]
let grover = lambda(string){
  let Uf = oracle(generateMatrix(f(string),4))
  let initialState = apply(tensor(Htwo,H),tensor([1 0 0 0],[0 1]))
  groverIter(initialState,Uf,4)
}
\end{lstlisting}
\end{center}

For example, running \texttt{grover} with input \texttt{[0 0 1 0]}, which amounts to searching for the binary string ``10'' gives us:
\begin{center}
\begin{lstlisting}[,mathescape=true]
Probability of state 0 is 0
Probability of state 1 is 0
Probability of state 2 is 0
Probability of state 3 is 0
Probability of state 4 is 0.5
Probability of state 5 is 0.5
Probability of state 6 is 0
Probability of state 7 is 0
\end{lstlisting}
\end{center}

This might look strange, after all shouldn't Grover give us the unique binary string s for which $f(s) = 1$ , with probability 1?
Instead what we get is that our system has a 50/50 chance of either being in the state $\ket{100}$ or the state $\ket{101}$. 
Fortunately our algorithm works as intended: indeed the ``top'' q-register is always in the state $\ket{10}$, but the ``bottom'' qubit was placed in a superposition earlier on, which results to our the collective system's probabilities being split 50/50 as above.

While it is easy to ``untangle'' the respective probabilities by hand for small systems, things can get considerably more complex in many-qubit systems.
To alleviate this, Qumin provides us with the \texttt{subsystems} function that given a state and a configuration, returns the respective probabilities for each subsystem/register.

Assuming we label our basis vectors with the usual binary indexing:\\

\begin{mdframed}
\noindent\textbf{Subsystems Algorithm.} \\
\textbf{Inputs:}  State: $\ket{\psi}, Configuration:  (c_1,c_2,...,c_n)$\\
\textbf{Outputs:} Measurement outcome probabilities per subsystem \\
\textbf{0:} Check $ log2(dim(H)) = \sum\limits_{i=1}^{n} c_n$, if not, exit with error.  \\
\textbf{1:} For each subsystem $S_i$, the probability of the state $\ket{s_{ik}}$, where $ 0 \leq k \leq c_i - 1$, is:   \\ \\
 $ \sum \frac{\braket{j|\psi} ^ 2}{|\ket{\psi}| ^ 2} $, or just $\sum \braket{j|\psi} ^ 2$ if we work with normalised states, for all basis vectors $\ket{j}$, that contain $k$ (in binary) as the $i$th  substring of their label.\\
 \end{mdframed}
We can then modify \texttt{groverRoutine} as such:

\begin{center}
\begin{lstlisting}[,mathescape=true]
let groverRoutine = lambda(state : qubit * qubit * qubit, iteration : !{operator[3]}, times : int){
  subsystems(applyN(iteration,state,times),[2 1])
}
\end{lstlisting}
\end{center}

The second argument of \texttt{subsystems}, namely \texttt{[2 1]}, tells the function to split our system into two subsystems: one of 2 qubits and another of 1.
Observe that there is no unique way to decompose this system, indeed any valid configuration may be supplied to \texttt{subsystems}. 
This means that a quantum system may be split into any configuration of subsystems of registers/qubits, if the sum of their dimensions equals the dimension of the system.

Finally, running \texttt{grover} again as above, yields:

\begin{center}
\begin{lstlisting}[,mathescape=true]
Probability of Subsystem0 state 00 is:  0
Probability of Subsystem0 state 01 is:  0
Probability of Subsystem0 state 10 is:  1.0
Probability of Subsystem0 state 11 is:  0
Probability of Subsystem1 state 0 is:  0.5
Probability of Subsystem1 state 1 is:  0.5
\end{lstlisting}
\end{center}

We can see that \texttt{subsystems}'s output is similar to that of \texttt{measurements}, but this time states are grouped by subsystem according to the configuration suplied and labeled in binary instead, to clearly present the specific states of the constituent subsystems of qubits.

\subsection{Generalising the implementation}

The program described in the previous section can easily be generalised, for example to work for databases of any size. We will now examine such a generalisation.

First we need need to create some helper functions, starting with one that generates the vector $\ket{0}$ for a space of given size, for use in initialising our program:

\begin{center}
\begin{lstlisting}[,mathescape=true]
let genZero = lambda(size){
  if((size = 1)){
    [1]
  }
  else{
    prepend(0,genZero((size - 1)))
  }
}
\end{lstlisting}
\end{center}

We also need a function that will compute the tensor product of an operator $U$, $n$ times, i.e. $U^{\otimes n}$:

\begin{center}
\begin{lstlisting}[,mathescape=true]
let tensorTimesN = lambda(op,n){
  if((n = 1)){
    op
  }
  else{
    (op $\otimes$ tensorTimesN(op,(n - 1)))
  }
}

\end{lstlisting}
\end{center}

We can now rewrite \texttt{grover} as:

\begin{center}
\begin{lstlisting}[,mathescape=true]
let grover = lambda(string){
  let N = length(string)
  let logN = logTwo . length(string)
  let Uf = oracle(generateMatrix(f(string),N))
  let initialState = apply(
                       tensor(tensorTimesN(H,logN),H),
                       tensor(genZero(N),[0 1]))
  groverIter(initialState,Uf,toInt(sqrt(N)),N)
}

\end{lstlisting}
\end{center}

and \texttt{groverIter} as:
\begin{center}
\begin{lstlisting}[,mathescape=true]
let groverIter = lambda(state,Uf,times,N){
  
  let hN = tensorTimesN(H,logTwo(N))
  
  let average = apply(hN, genZero(N))

  let groverOper =  ((2 * outer(average,average)) - generateMatrix(identity,length(average)))

  let opb = tensor(groverOper,generateMatrix(identity,2))

  let iteration = apply(opb,Uf)
  
  groverRoutine(state,iteration,times)
}
\end{lstlisting}
\end{center}

The  routine \texttt{groverRoutine} stays the same, except for a change in the argument types and the \texttt{subsystems} configuration argument, which must be edited to match the dimensions of the instance we are to execute.
For example for a 3-dimensional database we have:

\begin{center}
\begin{lstlisting}[,mathescape=true]
let groverRoutine = lambda(state : qubit * qubit * qubit * qubit, iteration : !{operator[4]}, times : int){
  subsystems(applyN(iteration,state,times),[3 1])
}
\end{lstlisting}
\end{center}

Finally, let's run our new generalised version of Grover's algorithm, for :
\[
 f(x) =
  \begin{cases} 
      \hfill 1    \hfill & \text{ if $x$ is ``000''} \\
      \hfill 0 \hfill & \text{ otherwise} \\
  \end{cases}
\]

\begin{center}
\begin{lstlisting}[,mathescape=true]
=> grover([1 0 0 0 0 0 0 0])
Probability of Subsystem0 state 000 is:  1.0
Probability of Subsystem0 state 001 is:  0.0
Probability of Subsystem0 state 010 is:  0.0
Probability of Subsystem0 state 011 is:  0.0
Probability of Subsystem0 state 100 is:  0.0
Probability of Subsystem0 state 101 is:  0.0
Probability of Subsystem0 state 110 is:  0.0
Probability of Subsystem0 state 111 is:  0.0
Probability of Subsystem1 state 0 is:  0.5
Probability of Subsystem1 state 1 is:  0.5
\end{lstlisting}
\end{center}

Which is what we expected to see.

\section{Conclusion and future directions}

As we saw, Qumin's features are centered around the idea of providing a simple framework that explicitly handles the interaction of quantum and classical computation and allows us to experiment with, and implement, algorithms, with an aim for simple and high-level programming.
The use of linear types guarantees us correctness in the quantum parts of our system, while the untyped part of the language allows us to make unrestricted use of familiar programming techniques for the classical part. The various programming techniques and patterns, such as matrix/oracle generators and measurement functions, allow us to clearly and easily implement various standard quantum algorithms and observe the results of their execution.

The main part to consider expanding is Qumin's type system. A major direction for such expansion would be the integration of dependent types \cite{cervesato1996linear,vakar2014syntax} which would greatly expand the expressive power of the quantum fragment of Qumin. For example in the generalised implementation of Grover's algorithm from last chapter, we
had to manually edit the type signature of our function to match the dimension of our database. While this might be a minor inconvenience here, it would be much cleaner to encode this dependency in the types themselves.

Such expansions, apart from allowing greater expressiveness when programming, are also interesting from a more theoretical viewpoint. For example, we would no longer need a ``signature schema'' for primitive functions, the dimension-dependency would be directly encoded in the signature of each function. This would also simplify the implementation of Qumin, which for now has ad hoc routines to handle primitive functions. 

Other interesting features include: a structure, like algebraic data types, that allows users to construct custom, complicated data structures and support for quantum control. The last one is especially interesting, as it would expand our language beyond the model of  ``classical control, quantum data''.

Yet another interesting area to consider are cross-compilation and/or interfacing of Qumin with current frameworks that aim to provide a generic instruction set or framework for quantum computation, as well as hardware-specific frameworks that correspond to actual implemented quantum computers. Such interfacing would allow Qumin's programs to be executed in experimental quantum computing devices and make use of various community provided optimisers, visualisers, circuit compilers etc. as described in aforementioned works.

\bibliography{references}{}
\bibliographystyle{plain} 
\newpage
\begin{appendices}
\section{Unitarity of \texttt{oracle}'s generated matrices}

\textbf{Proof.}
First we note that for the set of all binary strings of length $n$: $\{0,1\}^n $ and for a fixed string $c \in \{0,1\}^n$, the operation $f(x) : \{0,1\}^n \rightarrow \{0,1\}^n = x \oplus c$ is permutation on $\{0,1\}^n$. 
\begin{itemize}
\item $f$ is an injection: Let $s,v \in \{0,1\}^n$ and $f(s) = f(v) \implies s \oplus c = v \oplus c \implies s = v $ the last equality readily follows from the definition of bitwise exclusive or.
\item $f$ is a surjection: For any $s \in \{0,1\}^n$ we construct $v \in \{0,1\}^n$ such that $f(v) = s$. The construction is as follows: for the $i$th letter of $v$, we define the $i$th letter of $v$ as $v_i \oplus c_i = s_i$ where $v_i, s_i ,c_i$ are the $i$th letters of $v,s,c$.
\end{itemize}
We can now determine how $U(\ket{x,y}) = \ket{x, y \oplus f(x)}$ acts $V = X \otimes Y$ by observing how it acts on the usual orthonormal basis $b$ of $V$.
By fixing an element of $x_i \in X$ we get a permutation (as described above) on the following subset of our basis: $v_{x_i} = \{s \in b \ | \ s_i = x_i$ for $i = 0...n\}$, ie the set of basis vectors whose $n$ first digits match $x_i$.
This permutation induces a permutation operator $p_i$ on the subspace $span(v_{x_i})$. These subspaces do not overlap and their direct sum is the whole space $V$. Accordingly we can ``piece'' together the permutation operators to get a permutation operator on $V$ defined by
$p(s) = p_i$ for $v \in v_{x_i}$. The matrix representation of such operator is a block-diagonal permutation matrix, whose blocks would be the matrices of $p_i$s.

We can now show that permutation matrices are orthogonal:
\[ (PP^\dagger)_{ij} = \sum\limits_{k=1}^{n} P_{ik}P^\dagger_{kj} = \sum\limits_{k=1}^{n} P_{ik}P_{jk} \]
But we know that each column $k$ of P only contains a single non-zero entry of 1. So

\[ \sum\limits_{k=1}^{n} P_{ik}P{jk} = \begin{cases} 1  \ \mbox{if} \ i = j\\ 0 \ \mbox{if} \ i \neq j\end{cases}\]
which is the formula for the identity matrix, therefore $PP^\dagger = I$. Similarly we prove that $P^\dagger P = I$.

As a visual example, let us construct the operator $U(\ket{x,y}) = \ket{x, y \oplus f(x)}$ for the following function: \\

\begin{figure}[H]
  \begin{equation}
    f(x) =
    \begin{cases*}
      00 & if x = 00\\
      11 & if x = 01 \\
      10 & if x = 10 \\
      00 & if x = 11 \\
    \end{cases*}
  \end{equation}
\end{figure}

The corresponding permutations on the values of $y$ for fixed $x$ are:

\begin{table}[H]
\centering
\caption{for fixed x = 00}
\label{my-label1}
\begin{tabular}{l|l|l|l|l|l|}

$x,y:$ & 00,00 & 00,01 & 00,10 & 00,11 \\
$x,y\otimes f(x):$  & 00,00 & 00,01 & 00,10 & 00,11
\end{tabular}
\end{table}

\begin{table}[H]
\centering
\caption{for fixed x = 01}
\label{my-label2}
\begin{tabular}{l|l|l|l|l|l|}

$x,y:$ & 01,00 & 01,01 & 01,10 & 01,11 \\
$x,y\otimes f(x):$  &01,11 & 01,10 & 01,01 & 01,00
\end{tabular}
\end{table}

\begin{table}[H]
\centering
\caption{for fixed x = 10}
\label{my-label3}
\begin{tabular}{l|l|l|l|l|l|}

$x,y:$ & 10,00 & 10,01 & 10,10 & 10,11 \\
$x,y\otimes f(x):$  &10,10 & 10,11 & 10,00 & 10,01
\end{tabular}
\end{table}

\begin{table}[H]
\centering
\caption{for fixed x = 11}
\label{my-label4}
\begin{tabular}{l|l|l|l|l|l|}

$x,y:$ & 11,00 & 11,01 & 11,10 & 11,11 \\
$x,y\otimes f(x):$  &11,00 & 11,01 & 11,10 & 11,11
\end{tabular}
\end{table}

We observe that for each table, concatenating the comma-separated elements gives us an element of the standard orthonormal basis on the top row and the result of applying $U$ on it, on the second row. We can now list these strings in a orderly vertical fashion to get the matrix representation of $U$:\\

\begin{centering}
\begin{table}[H]
\center
    \begin{tabular}{|l|l|l|l|l|l|l|l|l|l|l|l|l|l|l|l|}
    \hline
   1 & ~ & ~ & ~ & ~ & ~ & ~ & ~ & ~ & ~ & ~ & ~ & ~ & ~ & ~ & ~ \\ \hline
   ~ & 1 & ~ & ~ & ~ & ~ & ~ & ~ & ~ & ~ & ~ & ~ & ~ & ~ & ~ & ~ \\ \hline
   ~ & ~ & 1 & ~ & ~ & ~ & ~ & ~ & ~ & ~ & ~ & ~ & ~ & ~ & ~ & ~ \\ \hline
   ~ & ~ & ~ & 1 & ~ & ~ & ~ & ~ & ~ & ~ & ~ & ~ & ~ & ~ & ~ & ~ \\ \hline
   ~ & ~ & ~ & ~ & ~ & ~ & ~ & 1 & ~ & ~ & ~ & ~ & ~ & ~ & ~ & ~ \\ \hline
   ~ & ~ & ~ & ~ & ~ & ~ & 1 & ~ & ~ & ~ & ~ & ~ & ~ & ~ & ~ & ~ \\ \hline
   ~ & ~ & ~ & ~ & ~ & 1 & ~ & ~ & ~ & ~ & ~ & ~ & ~ & ~ & ~ & ~ \\ \hline
   ~ & ~ & ~ & ~ & 1 & ~ & ~ & ~ & ~ & ~ & ~ & ~ & ~ & ~ & ~ & ~ \\ \hline
   ~ & ~ & ~ & ~ & ~ & ~ & ~ & ~ & ~ & ~ & 1 & ~ & ~ & ~ & ~ & ~ \\ \hline
   ~ & ~ & ~ & ~ & ~ & ~ & ~ & ~ & ~ & ~ & ~ & 1 & ~ & ~ & ~ & ~ \\ \hline
   ~ & ~ & ~ & ~ & ~ & ~ & ~ & ~ & 1 & ~ & ~ & ~ & ~ & ~ & ~ & ~ \\ \hline
   ~ & ~ & ~ & ~ & ~ & ~ & ~ & ~ & ~ & 1 & ~ & ~ & ~ & ~ & ~ & ~ \\ \hline
   ~ & ~ & ~ & ~ & ~ & ~ & ~ & ~ & ~ & ~ & ~ & ~ & 1 & ~ & ~ & ~ \\ \hline
   ~ & ~ & ~ & ~ & ~ & ~ & ~ & ~ & ~ & ~ & ~ & ~ & ~ & 1 & ~ & ~ \\ \hline
   ~ & ~ & ~ & ~ & ~ & ~ & ~ & ~ & ~ & ~ & ~ & ~ & ~ & ~ & 1 & ~ \\ \hline
   ~ & ~ & ~ & ~ & ~ & ~ & ~ & ~ & ~ & ~ & ~ & ~ & ~ & ~ & ~ & 1 \\ \hline
    \end{tabular}
\end{table}
\end{centering}

We can see now that each $x$ value defines a $n \times n$ block in our matrix, which contains a permutation of the corresponding $y$s. These can never overlap because their prefix varies by a constant value of $x$ and so we have a permutation matrix. $\blacksquare$

\section{Parsing expression grammar for Qumin}

Qumin's grammatical rules can be expressed in PEG as follows:

\noindent expr $\leftarrow$ \_ (load / func / ifelse / call / comp / infixCall / prefixCall/  list / assignment / boolLit / stringLit / complexLit / floatLit / intLit / name) \_ \\
numeral $\leftarrow$ ~``[0-9]+'' \\
call $\leftarrow$ name ``('' expr ((sep expr)*)? ``)'' \\
intLit $\leftarrow$ (``-'')? numeral \\
boolLit $\leftarrow$ \_(``\#t'' / ``\#f'')\_ \\
program $\leftarrow$ expr* \\
parameters $\leftarrow$ lvalue* \\
floatLit $\leftarrow$ (``-'')? numeral ``.'' numeral \\
stringLit $\leftarrow$ ````'' ~``[a-z A-Z 0-9 ! \# \$ ?]*'' ``'''' \\
load $\leftarrow$ ``--load'' \_ lvalue  \\
name $\leftarrow$ ~``[a-zA-Z⊗·+\-/*=?]+'' \\ 
comp $\leftarrow$ name (compsep name)+ \_ ``('' expr* ``)'' \\
complexLit $\leftarrow$ (``+''/``-'')? (numeral (``.'' numeral)?) (``+''/``-'') (numeral (``.'' numeral)?) ``i'' \\
list $\leftarrow$ ``['' expr* ``]'' \\
prefixCall $\leftarrow$ ``('' name expr ``)'' \\
compsep $\leftarrow$ \_``.''\_ \\
assignment $\leftarrow$ ``let'' \_ lvalue ``='' expr \\
ifelse $\leftarrow$ ``if'' \_ ``('' expr ``)'' ``\{'' expr* ``\}'' \_ ``else\'' \_ ``\{'' expr* ``\}''  \\
lvalue $\leftarrow$ ~"[a-zA-Z?]+" \_ \\
infixCall $\leftarrow$ ``('' expr name expr ``)'' \\
sep $\leftarrow$ \_ ``,'' \_  \\
func $\leftarrow$ ``lambda'' ``('' lvalue ((sep lvalue)*)? ``)'' ``\{'' expr* ``\}'' ( ``('' expr* ((sep expr)*)? ``)'' )? \\
\end{appendices}

\end{document}